\documentclass[9pt,twocolumn,twoside]{gsajnl}
\usepackage{epstopdf}
\usepackage{float}
\usepackage{textgreek}
\usepackage[section]{placeins}
\usepackage{gensymb}
\usepackage{siunitx}
\usepackage{tikz}
\usetikzlibrary{svg.path}
\definecolor{orcidlogocol}{HTML}{A6CE39}
\tikzset{
  orcidlogo/.pic={
    \fill[orcidlogocol] svg{M256,128c0,70.7-57.3,128-128,128C57.3,256,0,198.7,0,128C0,57.3,57.3,0,128,0C198.7,0,256,57.3,256,128z};
    \fill[white] svg{M86.3,186.2H70.9V79.1h15.4v48.4V186.2z}
                 svg{M108.9,79.1h41.6c39.6,0,57,28.3,57,53.6c0,27.5-21.5,53.6-56.8,53.6h-41.8V79.1z M124.3,172.4h24.5c34.9,0,42.9-26.5,42.9-39.7c0-21.5-13.7-39.7-43.7-39.7h-23.7V172.4z}
                 svg{M88.7,56.8c0,5.5-4.5,10.1-10.1,10.1c-5.6,0-10.1-4.6-10.1-10.1c0-5.6,4.5-10.1,10.1-10.1C84.2,46.7,88.7,51.3,88.7,56.8z};
  }
}

\newcommand\orcidicon[1]{\href{https://orcid.org/#1}{\mbox{\scalerel*{
\begin{tikzpicture}[yscale=-1,transform shape]
\pic{orcidlogo};
\end{tikzpicture}
}{|}}}}

\usepackage{hyperref}

\articletype{inv} % article type
\runningtitle{Proposal} % For use in the footer
\runningauthor{Jittprasong}

\title{A feasibility study proposal of the predictive model to enable the prediction of population susceptibility to COVID-19 by analysis of vaccine utilization for advising deployment of a booster dose}

\author[1,$\ast$]{Chottiwatt Jittprasong} 

\affil[1]{Biomedical Robotics Laboratory, Department of Biomedical Engineering, College of Engineering, City University of Hong Kong}

\correspondingauthoraffiliation[$\ast$]{Corresponding author: Biomedical Robotics Laboratory, Department of Biomedical Engineering, College of Engineering, City University of Hong Kong, Kowloon, Hong Kong.  \href{chottiwatt.j@my.cityu.edu.hk}{chottiwatt.j@my.cityu.edu.hk}}

\begin{abstract}
With the present highly infectious dominant SARS-CoV-2 strain of B1.1.529 or Omicron spreading around the globe, there is concern that the COVID-19 pandemic will not end soon and that it will be a race against time until a more contagious and virulent variant emerges. One of the most promising approaches for preventing virus propagation is to maintain continuous high vaccination efficacy among the population, thereby strengthening the population protective effect and preventing the majority of infection in the vaccinated population, as is known to occur with the Omicron variant frequently. Countries must structure vaccination programs in accordance with their populations' susceptibility to infection, optimizing vaccination efforts by delivering vaccines progressively enough to protect the majority of the population. We present a feasibility study proposal for maintaining optimal continuous vaccination by assessing the susceptible population, the decline of vaccine efficacy in the population, and advising booster dosage deployment to maintain the population's protective efficacy through the use of a predictive model. Numerous studies have been conducted in the direction of analyzing vaccine utilization; however, very little study has been conducted to substantiate the optimal deployment of booster dosage vaccination with the help of a predictive model based on machine learning algorithms.
\end{abstract}


\dates{\rec{25 04, 2022} \acc{25 04, 2022}}

\begin{document}

\maketitle
\thispagestyle{firststyle}
%\slugnote
%\firstpagefootnote
\vspace{-13pt}% Only used for adjusting extra space in the left column of the first page

\section{Introduction}

Since the outbreak of severe acute respiratory syndrome coronavirus 2 (SARS-CoV-2), the causal agent of Coronavirus Disease (COVID-19), the pandemic has wreaked havoc on humans and posed a grave danger to the world's economic growth. The virus spreads through close contact with an infected person, usually through respiratory droplets produced when an infected person coughs or sneezes. It has spread to more than 200 countries and territories, infecting more than 500 million people and killing more than 6 million people all over the world \citep{RN124}. However, the COVID-19 pandemic also demonstrated the superior capability of modern medicine, which was able to develop the first experimental vaccine 42 days after the disease emerged \citep{moderna}. Tozinamaren (BNT162b2), also known as the Pfizer–BioNTech COVID-19 vaccine sold under the brand name Comirnarty, was the first vaccine to be added to the World Health Organization's (WHO) Emergency Use List (EUL) on 31 December 2020, just a year after the first outbreak \citep{RN128,RN129}.
 
Numerous vaccinations are now being used worldwide, with some countries currently administering the fourth dose to their populations \citep{RN130,RN124} Currently, the SARS-CoV-2 variant of B.1.1.529 (Omicron) is responsible for the majority of infections occurring at present. Although less fatal, it is more infectious than the last dominant strain of B.1.617.2 (Delta). Moreover, it can also evade double vaccination and the immune system \citep{RN131}. Omicron, therefore, caused numerous reinfection and outbreak among the fully vaccinated individual. Those who have received a vaccine with less efficacy are more vulnerable to Omicron infection \citep{RN132}.
 
This raises concern that the COVID-19 pandemic may not be over anytime soon and that it is a race against time until a more resistant variant emerges. In the meanwhile, the most effective strategy to prevent the virus from spreading is to maintain continuous immunization, which increases the protective effect on the individual. This is because the vaccine's efficacy against the Omicron variant declines fairly quickly \citep{RN132}.
 
However, keeping up with the number of populations who received the first dose, second dose, third dose, fourth dose, and so on are time extensive considering the nature of the deployment of the vaccine, which focuses mainly on how quickly the vaccine can be deployed as much as possible not how can we track it afterward. In this paper, we propose a predictive model for predicting the susceptible vaccinated population using generalized vaccination distribution data, daily vaccination rate, and vaccine type, in conjunction with a recurrent neural network for predicting the spread of infection in a specific geographic area. 

This enables countries to plan their vaccination programs in accordance with their populations' susceptibility to infection, optimizing vaccination program by not deploying vaccines too rapidly but enough to protect the majority of the population from infection, thereby containing the spread of infection. 

\section{Proposal}
\label{sec:proposal}

To ascertain the population's susceptibility to COVID-19, we must first ascertain vaccination utilization in the region. The vaccination utilization data may comprise the vaccination rate, the proportion of types of vaccines deployed, the ordinal number of vaccine doses administered, and the vaccination population's age distribution. Additionally, the inclusion of statistics on the unvaccinated and partially vaccinated population will result in an increase in the prediction's accuracy.

\FloatBarrier
\begin{figure}[H]
\includegraphics[width=\linewidth]{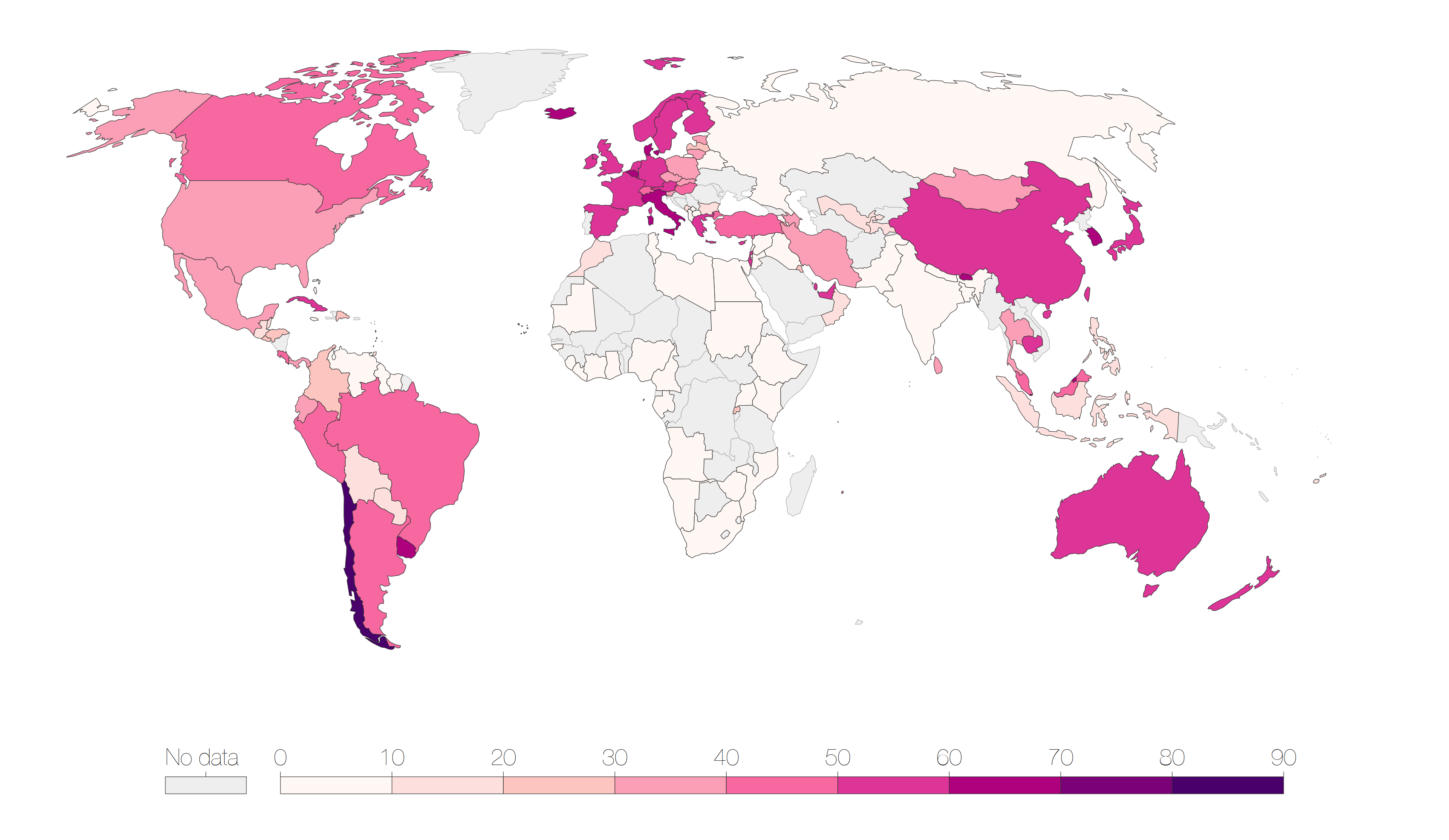}
\caption{A diagram illustrating COVID-19 vaccine boosters administered per 100 people in every country.\\
Note: From "A global database of COVID-19 vaccinations," by \cite{RN124}, \textit{Nature Human Behaviour}, 5, 947-953. Copyright 2022 by Our World in Data under CC BY 4.0 license. Reprinted under license terms.}
\label{fig:1}
\end{figure}
\FloatBarrier

As far as we are aware, the application of ANNs to predict vaccine/pharmaceutical utilization is extremely limited. However, the potential for machine learning to assist us in predicting population susceptibility to COVID-19 and optimizing vaccination programs is indeed intriguing and may benefit us for eradication of COVID-19 pandemic.  

The prediction model will let us see how effectively a country's population is protected against COVID-19 and when a booster dosage is required to maintain adequate protection. Due to the fact that each person got their vaccination doses at a different interval, determining when the population will need the booster dosage is challenging, if not impossible. As illustrated in \hyperref[fig:1]{Figure 1}, booster doses above the recommended double doses of immunizations are not distributed equally around the globe, as booster doses are not yet generally recommended in the majority of the world. 
 
The predictive model may be built on machine learning (ML) algorithms such as support vector machines, random forests, and k-nearest neighbors. In addition, the predictive model may also be built on statistical models such as logistic regression, generalized linear models, and generalized additive models.

\begin{figure}[H]
\includegraphics[width=\linewidth]{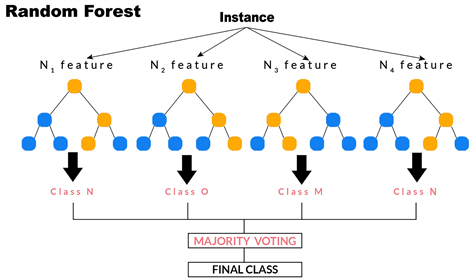}
\caption{A diagram depicting simplified principle of random forest machine learning algorithm.\\
Note: From "Effective Electrical Submersible Pump Management Using Machine Learning," by \cite{RN123}, \textit{Open Journal of Civil Engineering}, 11, 70-80. Copyright 2022 by Pham et al. and Scientific Research Publishing Inc. under CC BY-NC 4.0 license. Reprinted under license terms.}
\label{fig:2}
\end{figure}

\cite{RN120} observed that random forest \hyperref[fig:2]{(Figure 2)} was the most effective ML algorithm for predicting vaccine utilization. Additionally, the authors noted that the model outperformed the conventional technique of vaccination utilization tracking. Another study used the artificial neural network (ANN) to forecast pharmaceutical utilization based on geographical location \citep{RN121}.

\FloatBarrier
\FloatBarrier

\cite{RN125} have proposed numerous strategies for combating the COVID-19 pandemic through employing various machine learning algorithms, including the Recurrent Neural Network (RNN), the Long Short Term Memory (LSTM), the Generative Adversarial Network (GAN), and the Extreme Learning Machine (ELM). These algorithms may be used for a variety of purposes, including infection transmission prediction, medication dosage estimation, visualization of spread, and treatment recommendation. 

\cite{RN126} has developed a deterministic and stochastic RNN of LSTM and Mixture Density Network (MDN) for real-time prediction of the virus accross United States. 

\FloatBarrier
\begin{figure}[H]
\includegraphics[width=\linewidth]{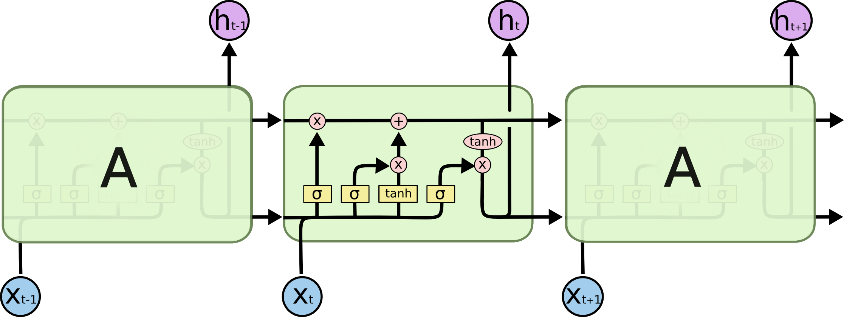}
\caption{A diagram illustrating the structure of long short term memory recurrent neural network.\\
Note: From "Optimizing COVID-19 vaccine distribution across the United States using deterministic and stochastic recurrent neural networks," by \cite{RN126}, \textit{PLoS ONE }, 16(7): e0253925. Copyright 2021 by Davahli et al. under CC BY 4.0 license. Reprinted under license terms.}
\label{fig:3}
\end{figure}
\FloatBarrier

By calculating the mixing coefficient, mean, and standard deviation (SD), the MDNs enables the estimation of the mixed distribution. This way, MDNs estimate probability distributions for potential outcomes rather than producing completely defined outputs like what LSTMs do. 
 
\FloatBarrier
\begin{figure}[H]
\includegraphics[width=\linewidth]{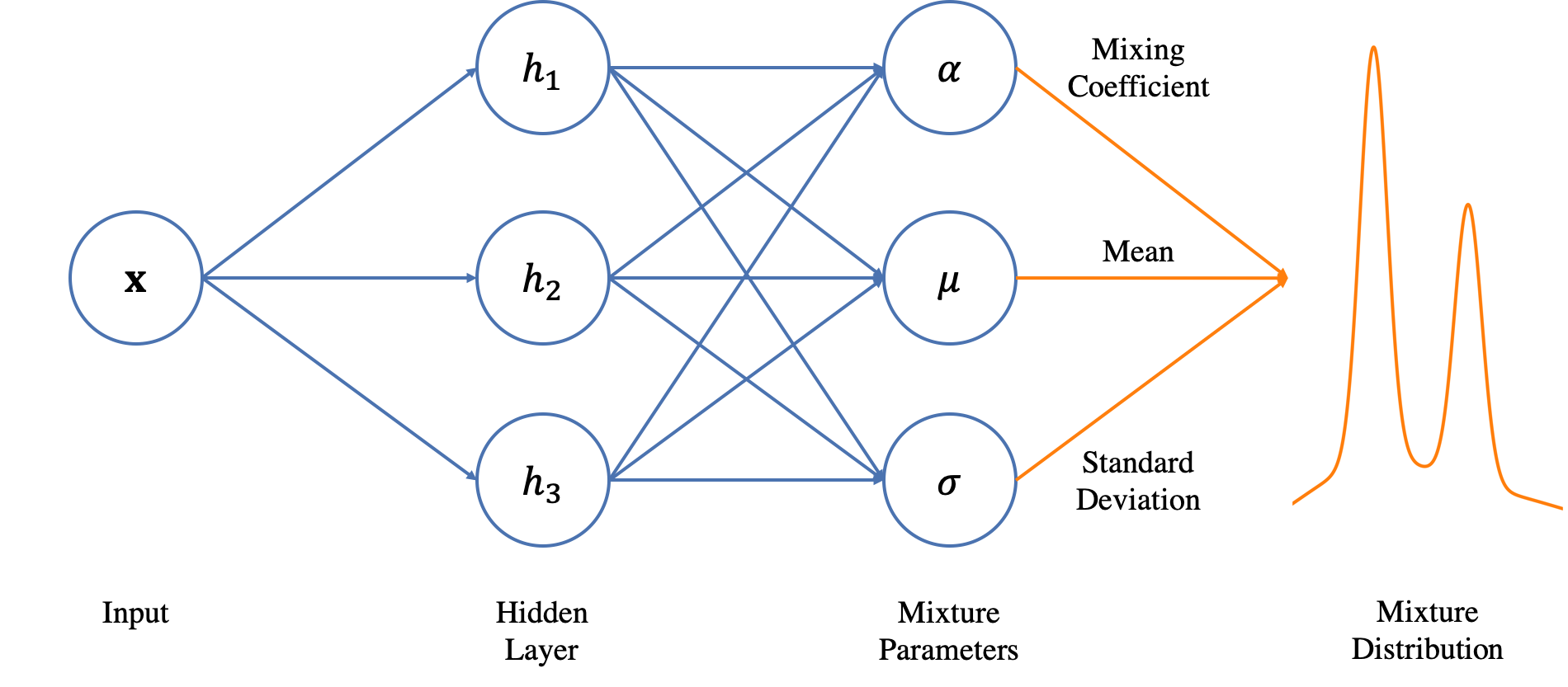}
\caption{A diagram illustrating the structure of mixed desity network.\\
Note: From "Optimizing COVID-19 vaccine distribution across the United States using deterministic and stochastic recurrent neural networks," by \cite{RN126}, \textit{PLoS ONE }, 16(7): e0253925. Copyright 2021 by Davahli et al. under CC BY 4.0 license. Reprinted under license terms.}
\label{fig:4}
\end{figure}
\FloatBarrier

By combining LSTM and MDN, the COVID-19 spread of infection predictive model is developed.
We can reliably predict the next outbreak and spread of infection using two sequence-learning models: a deterministic LSTM model that delivers deterministic output and a stochastic LSTM/MDN model.
Both the deterministic LSTM model and the stochastic LSTM/MDN model performed well in predicting the infection trend. 

\FloatBarrier
\begin{figure}[H]
\includegraphics[width=\linewidth]{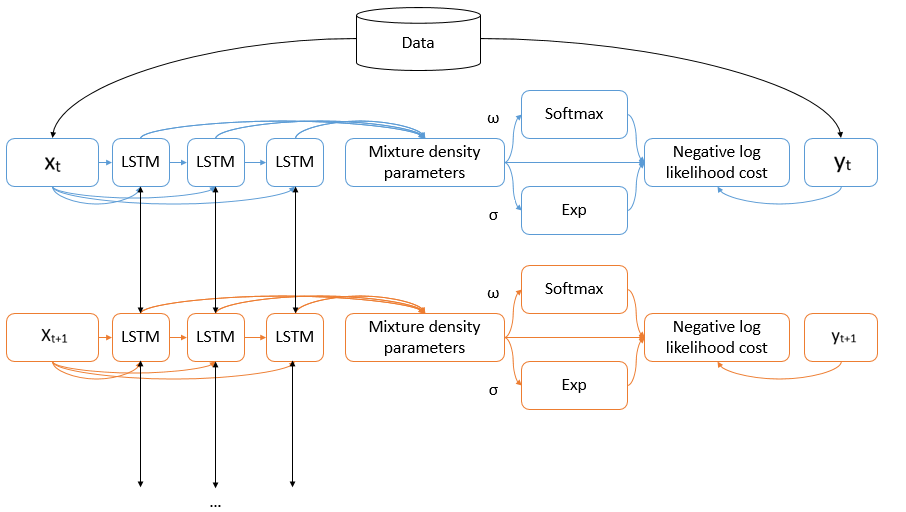}
\caption{A diagram illustrating the structure of stochastic LSTM/MDN model.\\
Note: From "Optimizing COVID-19 vaccine distribution across the United States using deterministic and stochastic recurrent neural networks," by \cite{RN126}, \textit{PLoS ONE }, 16(7): e0253925. Copyright 2021 by Davahli et al. under CC BY 4.0 license. Reprinted under license terms.}
\label{fig:4}
\end{figure}
\FloatBarrier

This demonstrated that machine learning may be repurposed from predicting infection transmission to predicting optimal vaccine distribution, since both of these characteristics are critical to the success of disease control, where viral transmission is continuously monitored and susceptible populations are protected. 

Infact, one possible strategy has been proposed, but it focuses only on vaccination distribution without booster doses and excludes the decline in vaccine efficacy and virus reproduction number.  Although not yet implemented, \cite{VacSim} has built a pipeline called VacSIM for optimizing COVID-19 vaccine distribution using Deep Reinforcement Learning models and contextual bandit machine learning framework.

\FloatBarrier
\begin{figure}[H]
\includegraphics[width=\linewidth]{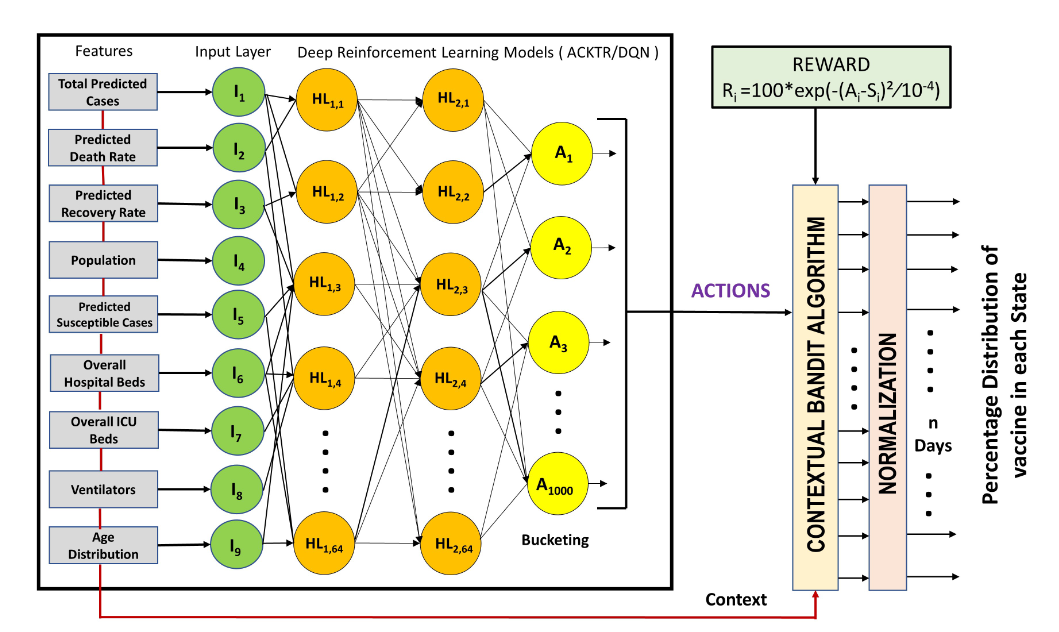}
\caption{A diagram showing the structure of VacSIM deep reinforcement learning models.\\
Note: From "VacSIM: Learning Effective Strategies for COVID-19 Vaccine Distribution using Reinforcement Learning," by \cite{VacSim}, \textit{arXiv}. Copyright 2021 by Awasthi et al. under CC BY-NC-SA 4.0 license. Reprinted under license terms.}
\label{fig:4}
\end{figure}
\FloatBarrier

Another study developed a pandemic vulnerability index (PVI) dashboard using machine learning, statistical modeling, and visualization to monitor the susceptibility of each region's population currently operating under the National Institute of Environmental Health Sciences in the United States only \citep{RN127}.

\FloatBarrier
\begin{figure}[H]
\includegraphics[width=\linewidth]{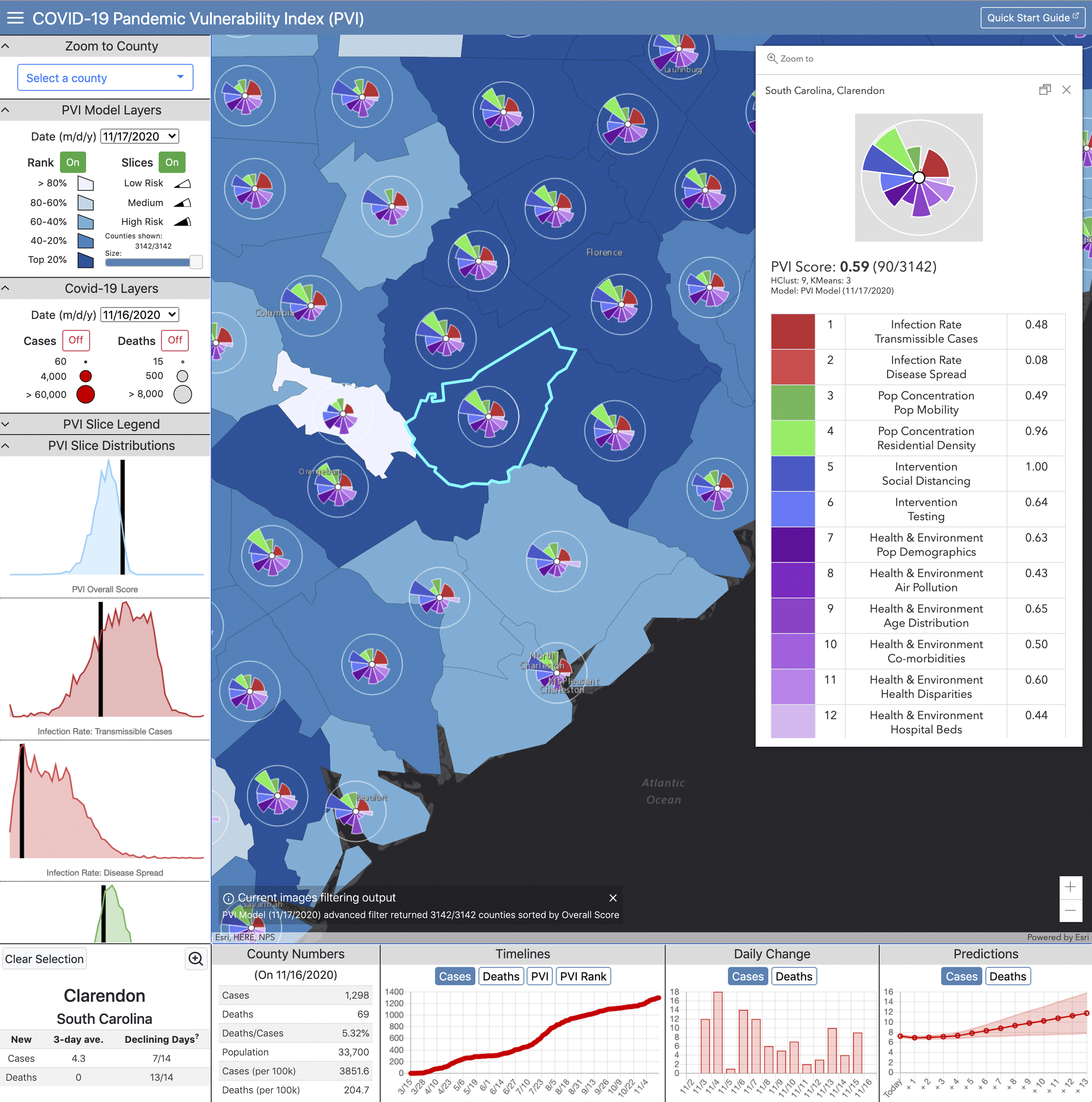}
\caption{A diagram depicting the Pandemic Vulnerability Index in action.\\
Note: From "The COVID-19 Pandemic Vulnerability Index (PVI) Dashboard: Monitoring County-Level Vulnerability Using Visualization, Statistical Modeling, and Machine Learning," by \cite{RN127}, \textit{Environmental Health Perspectives}. Reprinted from Environmental Helth Perspectives as a public domain.}
\label{fig:4}
\end{figure}
\FloatBarrier

Although the application is primarily for the United States and the PVI does not taken to account vaccinated population, it demonstrates that a predictive model based on machine learning and statistical models is capable of tracking vaccine booster doses and recommending an ideal vaccination scheme. 

Despite the fact that numerous studies have been conducted in a similar direction of developing a predictive model for predicting vaccine efficacy decline over time and population susceptibility, it would require state-of-the-art data analysis as well as robust machine learning algorithms to carefully analyze vaccination data from the time the first vaccine was administered to the present and in real-time. This would be a difficult task, but one that would be very beneficial to the public health community and the fight against COVID-19.

\section{Implications}
The development of a predictive model capable of predicting vaccine efficacy decline in the population and advising on when to initiate a booster dose vaccination program to maintain an adequate level of protection in the community would be required to contain the pandemic for an extended period of time until it subsides. This would need the development of a model capable of predicting vaccine efficacy in the population over time. Additionally, the model must be capable of predicting the effect of vaccination on disease transmission within the population.

If successful, we will have control over the pandemic's direction and will have an upper hand containing the virus by keeping the population adequately protected against COVID-19.

\section{Data availability}

Data sharing not applicable to this report as no datasets were generated or analysed during the current study.

\section{Funding}
This article received no specific grant from any funding agency in the public, commercial, or not-for-profit sectors.
\section{Conflicts of interest}
The author reports no conflict of interest.

\bibliography{bibliography}

\end{document}